\documentclass[letter]{emulateapj}

\usepackage{natbib}
\usepackage{amssymb}
\usepackage{amsmath}
\usepackage{color}

\newcommand{\uKamin}{$\mu$K$_{\textrm{CMB}}$-arcmin}

\begin{document}

\title{The Contribution of Radio Galaxy Contamination to Measurements of
  the Sunyaev-Zel'dovich Decrement in Massive Galaxy Clusters at 140~GHz with Bolocam}

\author{J.~Sayers\altaffilmark{1,8},
   T.~Mroczkowski\altaffilmark{1,2,3},
   N.~G.~Czakon\altaffilmark{1},
   S.~R.~Golwala\altaffilmark{1},
   A.~Mantz\altaffilmark{4},
   S.~Ameglio\altaffilmark{5},
   T.~P.~Downes\altaffilmark{1},
   P.~M.~Koch\altaffilmark{6},
   K.-Y.~Lin\altaffilmark{6},
   S.~M.~Molnar\altaffilmark{7},
   L.~Moustakas\altaffilmark{2},
   S.~J.~C.~Muchovej\altaffilmark{1},
   E.~Pierpaoli\altaffilmark{5},
   J.~A.~Shitanishi\altaffilmark{5},
   S.~Siegel\altaffilmark{1},
   \& K.~Umetsu\altaffilmark{6}
 }
\altaffiltext{1}
  {Division of Physics, Math, and Astronomy, California Institute of Technology, Pasadena, CA 91125}
\altaffiltext{2}
  {Jet Propulsion Laboratory, Pasadena, CA 91109}
\altaffiltext{3}
  {NASA Einstein Postdoctoral Fellow}
\altaffiltext{4}
  {Kavli Institute for Cosmological Physics, University of Chicago, 5640 South Ellis Avenue, Chicago, IL 60637}
\altaffiltext{5}
  {University of Southern California, Los Angeles, CA 90089}
\altaffiltext{6}
  {Institute of Astronomy and Astrophysics, Academia Sinica, 
    P.O. Box 23-141, Taipei 10617, Taiwan}
\altaffiltext{7}
  {LeCosPA Center, National Taiwan University, 
    Taipei 10617, Taiwan}
\altaffiltext{8}
  {jack@caltech.edu}

\begin{abstract}
We describe in detail our characterization of the compact radio source population
in 140~GHz Bolocam observations of a set of 45 massive galaxy clusters.
We use a combination of 1.4 and 30~GHz data to select a total of 28 
probable cluster member
radio galaxies and also to predict their 140~GHz flux densities.
All of these galaxies are steep-spectrum
radio sources and they are found preferentially in the cool-core clusters
within our sample.
In particular, 11 of the 12 brightest cluster member 
radio sources are associated with
cool-core systems.
Although none of the individual galaxies are robustly detected in the 
Bolocam data, the ensemble-average flux density at 140~GHz
is consistent with, but slightly lower than,
the extrapolation from lower frequencies
assuming a constant spectral index.
In addition, 
our data indicate an intrinsic scatter of $\simeq 30$\% around the
power-law extrapolated flux densities at 140~GHz, although
our data do not tightly constrain this scatter.
For our cluster sample, which is composed of high mass and moderate redshift systems,
we find that the maximum fractional change in the Sunyaev-Zel'dovich signal 
integrated over any single
cluster due to the
presence of these radio sources is $\simeq 20$\%, and only $\simeq 1/4$ of
the clusters show a fractional change of more than 1\%.
The amount of contamination is strongly dependent on cluster morphology,
and nearly all of the clusters with $\ge 1$\% contamination
are cool-core systems.
This result indicates that radio contamination is not significant compared
to current noise levels in
140~GHz images of massive clusters and is in good agreement 
with the level of radio contamination
found in previous results based on lower frequency data or simulations.

\end{abstract}

\section{Introduction}

Recently, large Sunyaev-Zel'dovich (SZ) effect surveys 
from the Atacama Cosmology Telescope (ACT), the South
Pole Telescope (SPT), and \emph{Planck} have
delivered catalogs with hundreds of massive galaxy clusters
\citep{sunyaev72, vanderlinde10, marriage11a, reichardt12, planck11_viii}.
In the case of ACT and SPT, which operate from the ground
with modest frequency coverage near 150~GHz and arcminute 
resolution, unresolved galaxies present a potential
systematic uncertainty in characterizing (or even
detecting) galaxy clusters via the SZ effect \citep{cooray98, massardi04, coble07, lin09, sehgal10}.
Both instruments have accurately measured
the blank sky population of bright radio sources 
\citep{vieira10, marriage11}
and the anisotropy power spectrum from both radio and 
dusty submillimeter sources \citep{das11, reichardt12b}.
However, there are few observational constraints on the 150~GHz
emission from member
galaxies in massive clusters, even though clusters are known 
to host a significant number of radio galaxies.
For example, within 0.5~arcmin of the cluster center there are $\simeq 30$
times more radio sources compared to blank sky and
from 0.5 to $\simeq 2$~arcmin of the cluster center there
are $\simeq 3$ times more radio sources compared to 
blank sky \citep{coble07, muchovej10}.
Unfortunately, the study of extragalactic radio sources at mm
wavelengths is relatively immature,
and it is therefore non-trivial to properly account for
these radio sources in 150~GHz SZ data. 
Even worse, the bulk of the 150~GHz observational data is limited 
to very bright flat-spectrum sources and is
not particularly relevant to the dimmer 
steep-spectrum sources typical
of central cluster galaxies
\citep{coble07, sadler08, vieira10, marriage11, gold11, sajina11, planck11_xiii}. 

Fortunately, the low frequency properties of radio
sources are well understood.
First, radio sources are generally grouped into two families:
flat-spectrum and steep-spectrum. Steep spectrum
sources have a $\simeq 1$~GHz spectral index of $\alpha<-0.5$
and flat spectrum sources have a $\simeq 1$~GHz spectral
index of $\alpha>-0.5$, where $\alpha$ describes a 
source spectrum of the form
$\nu^{\alpha}$ \citep{dezotti10}.
Physically, a flat spectrum is typically due to
synchrotron self-absorption (i.e., the medium is
optically thick), which usually means the jet
is oriented towards the viewer. In contrast, when the
jet is orthogonal to the viewing angle, the optically
thin lobes are visible, which tend to have a steep
spectrum \citep{dezotti10}. 
For both families of radio sources, the spectral index tends to
decrease at higher frequencies due to electron
aging (relativistic energy loss) and/or the medium becoming
less optically thick at higher frequencies
\citep{dezotti10, kellermann66}.
This steepening has been observed in both flat spectrum
sources (e.g., \citet{sadler08, marriage11, massardi10, planck11_xiii,sajina11})
and steep spectrum sources (e.g., \citet{ricci06, tucci11}),
although some measurements do not find such steepening
(e.g., \citet{lin09, vieira10}).
In addition, steep spectrum sources are generally associated with powerful
elliptical and S0 galaxies (i.e., the BCG of a massive cluster), 
and nearly all of the
radio sources associated with clusters have a steep
spectrum (e.g., \citet{tucci11, coble07}).
Finally, steep spectrum sources tend to have little
or no temporal variability 
\citep{tucci11, tingay03, sadler06, bolton06}.
This wealth of knowledge has been exploited in a range
of calculations and simulations to estimate the amount
of radio contamination in 150~GHz SZ measurements
\citep{lin09, sehgal10, andersson11}.
These results indicate that radio contamination in massive
clusters at 150~GHz should be relatively minor,
but they have not yet been systematically 
verified via observational data.

\section{Data Reduction}

\subsection{Bolocam}

During the period from 2006 to 2012 we observed a set of 45
massive galaxy clusters using Bolocam at the 
Caltech Submillimeter Observatory
\citep{glenn98, haig04, czakon12, sayers12_2}.
These clusters have a median mass of 
M$_{500} \simeq 9 \times 10^{14}$~M$_{\odot}$ and masses
as low as M$_{500} \simeq 3 \times 10^{14}$~M$_{\odot}$,
which is similar to the SPT and ACT mass limits
\citep{marriage11a, reichardt12}.\footnote{
  M$_{500}$ is the mass enclosed within a sphere with an
  average density of 500 times the background density.}
All of the masses were computed using \emph{Chandra} X-ray
data according to the methods described in \citet{mantz10_ii}.
The median redshift of our sample is $z = 0.42$, and the median SZ S/N is 12.
A complete description of these observations and data are given elsewhere
\citep{sayers11, czakon12, sayers12_2},
and we briefly summarize the relevant details below.
For all of these observations, Bolocam was configured to operate at
140~GHz,\footnote{
  The SZ emission-weighted band center of Bolocam is 140.0~GHz. The effective band
  center for typical radio source spectrum is slightly lower ($138.7 - 139.2$~GHz for 
  the radio sources with $-1.5 \le \alpha \le -0.5$). This difference
  in effective band center produces less than a 1\% change in the
  extrapolated flux density from 30~GHz, and is therefore not
  included in our analysis.
  Furthermore, we note that our data are calibrated against objects
  with approximately thermal spectrum, and both the source spectrum
  and the Bolocam spectral response were fully accounted for
  in the calibration \citep{sayers12}.}
with a 58~arcsec full-width at half-maximum (FWHM) point-spread
function (PSF).
All of the cluster images are $14 \times 14$~arcmin squares centered 
on the cluster.
The centers of the images generally have noise RMSs of $0.7-1.5$~mJy/beam,
($15-30$~\uKamin) increasing to approximately twice that value at the lower coverage
edges of the images.
This depth is similar to the SPT and ACT survey depths~\citep{marriage11a, reichardt12}.

In order to subtract atmospheric fluctuations from these data,
we remove the time-instantaneous average signal over the
field of view and then high pass filter the datastreams
at a characteristic frequency of 250~mHz.
This process also removes some astronomical signal from
the data, and we quantify this via simulation to
obtain a signal transfer function.
For the results presented in this manuscript, the
cluster signal transfer function was computed as
described in detail in \citet{sayers11}.
We did not explicitly compute transfer functions for the
point sources.
Instead, we processed point source models centered on
the galaxy positions determined from low frequency data
through our reduction
pipeline and then directly fit the resulting filtered image
to our data.
In all cases the point source 
models were normalized to an amplitude of 1~mJy.
We note that the effective filtering of the point
source model is independent of the amplitude of the
model for sources $\lesssim 100$~Jy, which is at
least 3 orders of magnitude brighter than any 
of the radio sources in our sample.

\subsection{OVRO/BIMA and SZA}

Most of our radio source flux densities at $\simeq 30$~GHz were obtained from 
previously published results from the Owens Valley Radio Observatory (OVRO),
the Berkeley-Illinois-Maryland Association (BIMA) array, and the Sunyaev-Zel'dovich
Array (SZA) \citep{coble07, bonamente11}.
However, we reduced data collected by these facilities for
an additional 13 radio sources (3 from OVRO/BIMA and 10 from SZA).
Our reduction of these previously unpublished data followed the methods described in
e.g., \citet{coble07} and \citet{muchovej07}, and we refer the reader
to those manuscripts for additional details.
To obtain point source flux densities we simultaneously fit 
for the cluster SZ signal and all of the point sources.
We modeled the cluster SZ emission using the best-fit generalized
Navarro, Frenk, and White (gNFW) pressure profiles from
\citet{arnaud10}, which are described by
\begin{equation}
      P(X) = \frac{P_0}{( C_{500} X )^{\mathcal{\gamma}} [1 + 
      ( C_{500} X )^\mathcal{\alpha} ]^{(\mathcal{\beta}
      -\mathcal{\gamma})/\mathcal{\alpha}}},
\end{equation}
where $P$ is the pressure, $P_0$ is the pressure normalization,
$X$ is the radial coordinate, 
$C_{500}$ is the concentration parameter, and 
$\alpha$, $\beta$, and $\gamma$ describe the power-law
slopes at moderate, large, and small radii.
Corrections for the SZA primary beam
were accounted for in both the cluster and point source fits.

\section{Cluster-Member Radio Galaxies}

\subsection{Central Radio Galaxies - NVSS Selection}

\begin{figure*}
  \includegraphics[height=0.35\textwidth]{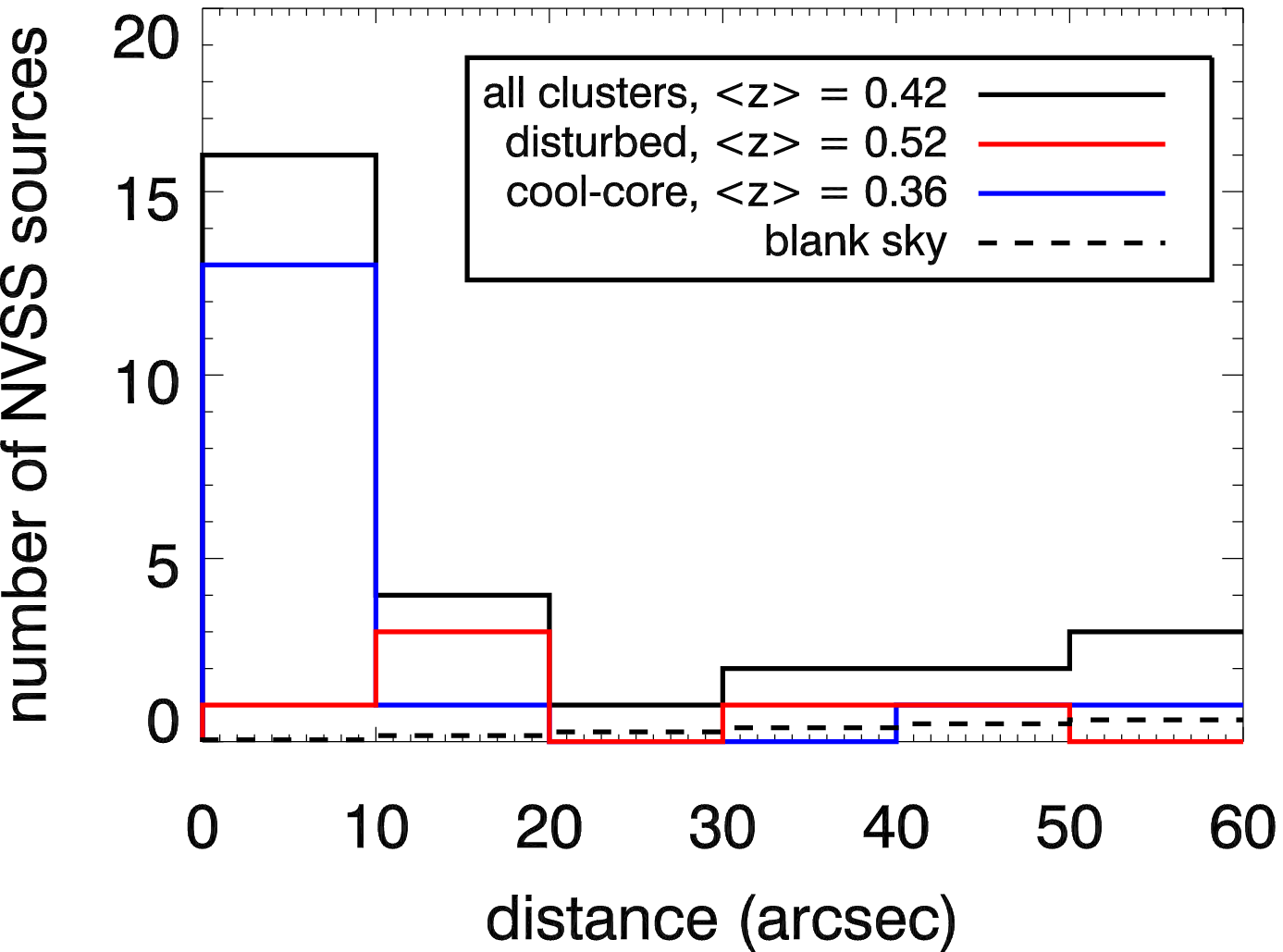}
  \hspace{0.05\textwidth}
  \includegraphics[height=0.35\textwidth]{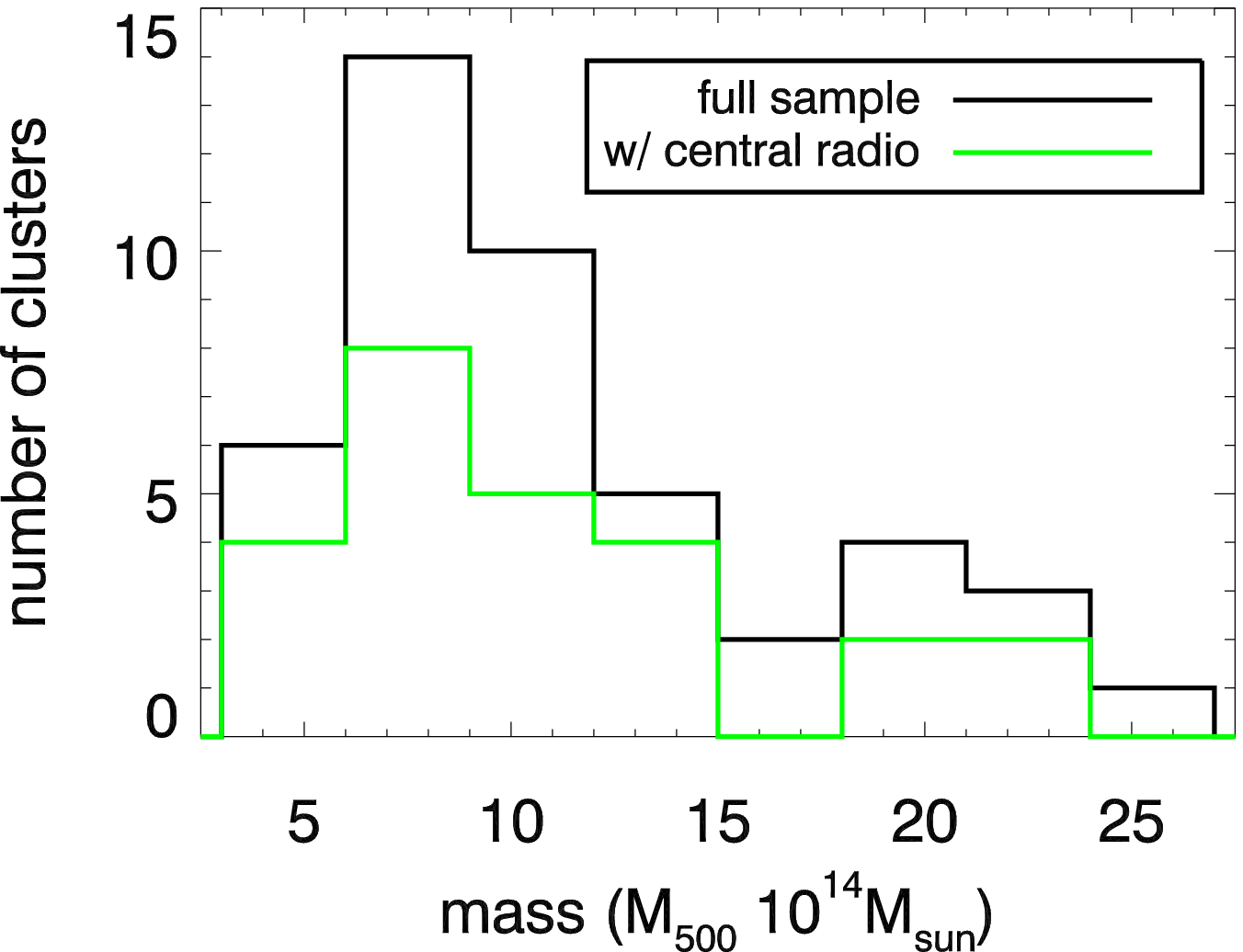}
  \caption{Left: the number of NVSS sources as a function of radius
    from the cluster centers. We show the number of NVSS sources
    for the full sample of 45 clusters, the 16 disturbed
    clusters, and the 17 cool-core clusters.
    Note the significant number of NVSS sources at the
    very center of cool-core clusters.
    For reference, we also show the blank-sky average
    number of sources for the NVSS survey given the
    angular size of each bin.
    Right: cluster masses for both the full
    sample of 45 objects and for the subset of 25 objects
    with central radio sources.
    The masses were computed from \emph{Chandra} X-ray data
    according to the methods described in \citet{mantz10_ii}.}
  \label{fig:number}
\end{figure*}

There are a significant number of bright radio galaxies near
the centers of massive clusters
(e.g., \citet{coble07} find that $\simeq 97$\%
of radio sources within 0.5~arcmin of the cluster center
are cluster members).
Unfortunately, these sources are difficult to detect in our
arcmin-resolution SZ data due to the large degeneracy between the inner slope
of the SZ profile and the radio source flux density.
Consequently, we have searched the 1.4~GHz NVSS catalog within 1~arcmin
of each cluster center \citep{condon98} in order to select
probable cluster member galaxies.
We chose the NVSS survey for our galaxy selection because
it has approximately uniform depth over the full
sky above dec $=-40$, and is complete above a flux density
of 2.5~mJy.
Throughout this paper we refer to the NVSS selected sources
within 1~arcmin as cluster member galaxies. 
However, based on the results of \citet{coble07} and the
blank-sky source density of the NVSS catalog, we
expect $\gtrsim 5$\% of the sources to be field 
galaxies unassocated with the clusters.

Our search radius of 1~arcmin was chosen for two main
reasons.
First, the results of \citet{coble07} indicate that a 1~arcmin 
search radius is large enough to contain the bulk of 
the bright cluster member radio galaxies
without including a significant number
of foreground and background galaxies that are not associated
with the cluster
(see Figure~\ref{fig:number}).
Second, by chosing a search radius approximately equal to
our beam FWHM we include all of the galaxies that won't
be resolved separately from the cluster SZ peak.
Our choice of a fixed angular search radius means that
the search radius in physical distance (e.g. kpc) varies
significantly over our sample, which could produce
biases in our galaxy selection.
However, given that the number of NVSS sources falls dramatically
beyond $\simeq 10$~arcsec in radius, it is unlikely that
a significant number of bright cluster member galaxies will
fall outside of our search radius.
Furthermore, as justified above, our search radius is small
enough such that the contamination from non-cluster-galaxies
is expected to be small.
Therefore, we do not expect any significant biases in our 
galaxy selection due to our choice of search radius.

Over our sample of 45 clusters, we find 31 NVSS sources spread
between 26 clusters
(three clusters contain two sources and one cluster contains three
sources, see Table~\ref{tab:radio_1}).
In agreement with previous results (e.g., \citet{mittal09}),
most of our cool-core clusters have at least one compact NVSS source
(15/17), while less than half of the disturbed clusters
in our sample have a compact NVSS source (6/16).
Note that our classification of clusters as cool-core or disturbed
is based on X-ray measurements and is described in detail
in \citet{sayers12_2}.
Briefly, the projected luminosity ratio 
($\textrm{L}_{\textrm{rat}} = \textrm{L}(\textrm{R} < 0.05\textrm{R}_{500}) / 
\textrm{L}(\textrm{R} < \textrm{R}_{500})$)
is used to define
cool-core systems \citep{mantz09, mantz10_ii, bohringer10},
with clusters having L$_{\textrm{rat}} >0.17$ classified
as cool-cores. 
The centroid shift parameter (w$_{500}$), which measures the difference
between the X-ray peak and the X-ray centroid as a function of radius,
is used to define disturbed systems \citep{maughan08, pratt09, maughan12},
with clusters having w$_{500} \ge 0.01$ classified as disturbed.
We note that two systems are both cool-core and disturbed,
so the number of non-cool-core disturbed systems with
compact NVSS sources is 4/14.

\begin{deluxetable*}{cccc|cccc} 
  \tabletypesize{\tiny}
  \tablewidth{\textwidth}
   \tablecaption{NVSS sources within 1 arcmin of the cluster center}
   \tablehead{\colhead{cluster} & \colhead{redshift} & \multicolumn{2}{c}{morphology} & 
     \multicolumn{4}{c}{NVSS sources} \\ 
     \colhead{} & \colhead{} & \colhead{cool-core} & \colhead{disturbed} & 
     \colhead{RA (J2000)} & \colhead{dec (J2000)} & \colhead{distance (arcsec)} & \colhead{1.4 GHz flux density (mJy)}}
   \startdata
    MACS J0018.5 & 0.54 & & & \multicolumn{4}{c}{no sources} \\
    MACS J0025.4 & 0.58 & & & 00:25:32.0 & $-$12:23:09.1 & 39 & $\phn28.7 \pm 1.3^{\phm{\star}}$ \\
    ZWCL 0024 & 0.39 & & $\checkmark$ & 00:26:35.5 & +17:09:31.7 & 10 & $\phn\phn2.6 \pm 0.4^{\phm{\star}}$ \\
    Abell 209 & 0.21 & & & 01:31:52.7 & $-$13:36:59.6 & 13 & $\phn18.4 \pm 1.0^{\phm{\star}}$ \\
    CL J0152.7 & 0.83 & & $\checkmark$ & \multicolumn{4}{c}{no sources} \\
    Abell 267 & 0.23 & & $\checkmark$ & \multicolumn{4}{c}{no sources} \\
    Abell 370 & 0.38 & & $\checkmark$ & 02:39:55.3 & $-$01:34:18.3 & 37 & $\phn10.5 \pm 1.0^{\phm{\star}}$ \\
    Abell 383 & 0.19 & $\checkmark$ & & 02:48:03.4 & $-$03:31:43.9 & $\phn3$ & $\phn40.9 \pm 1.3^{\phm{\star}}$ \\
    MACS J0257.1 & 0.50 & & & \multicolumn{4}{c}{no sources} \\
    MACS J0329.6 & 0.45 & $\checkmark$ & $\checkmark$ & 03:29:41.7 & $-$02:11:52.2 & $\phn7$ & $\phn\phn6.9 \pm 0.6^{\phm{\star}}$ \\
    MACS J0416.1 & 0.42 & & $\checkmark$ & \multicolumn{4}{c}{no sources} \\
    MACS J0417.5 & 0.44 & $\checkmark$ & $\checkmark$ & 04:17:34.9 & $-$11:54:34.2 & 12 & $\phn29.8 \pm 1.7^{\phm{\star}}$ \\
    MACS J0429.6 & 0.40 & $\checkmark$ & & 04:29:36.0 & $-$02:53:06.4 & $\phn0$ & $138.8 \pm 4.2^{\phm{\star}}$ \\
    MACS J0451.9 & 0.43 & & $\checkmark$ & \multicolumn{4}{c}{no sources} \\
    MACS J0454.1 & 0.55 & & $\checkmark$ & \multicolumn{4}{c}{no sources} \\
    MACS J0647.7 & 0.59 & & & \multicolumn{4}{c}{no sources} \\
    MACS J0717.5 & 0.55 & & $\checkmark$ & 07:17:34.1 & +37:45:01.3 & 31 & $\phn90.9 \pm 3.7^{\phm{\star}}$ \\
                 &      & &              & 07:17:35.7 & +37:45:39.2 & 47 & $102.5 \pm 3.7^{\phm{\star}}$ \\
    MACS J0744.8 & 0.69 & & $\checkmark$ & \multicolumn{4}{c}{no sources} \\
    Abell 611 & 0.29 & & & \multicolumn{4}{c}{no sources} \\
    Abell 697 & 0.28 & & & \multicolumn{4}{c}{no sources} \\
    MACS J0911.2 & 0.50 & & & \multicolumn{4}{c}{no sources} \\
    Abell 963 & 0.21 & & & \multicolumn{4}{c}{no sources} \\
    MS 1054 & 0.83 & & $\checkmark$ & 10:56:59.6 & $-$03:37:26.8 & 18 & $\phn14.1 \pm 0.9^{\phm{\star}}$ \\
    MACS J1115.8 & 0.36 & $\checkmark$ & & 11:15:51.8 & +01:29:55.5 & $\phn2$ & $\phn16.2 \pm 1.0^{\phm{\star}}$ \\
    MACS J1149.5 & 0.54 & & $\checkmark$ & \multicolumn{4}{c}{no sources} \\
    Abell 1423 & 0.21 & & & 11:57:16.8 & +33:36:44.6 & $\phn9$ & $\phn33.3 \pm 0.9^{\phm{\star}}$ \\
    MACS J1206.2 & 0.44 & & & 12:06:12.1 & $-$08:48:02.5 & $\phn4$ & $160.9 \pm 6.3^{\phm{\star}}$ \\
    CL J1226 & 0.89 & & & 12:26:58.3 & +33:32:44.3 & $\phn7$ & $\phn\phn4.3 \pm 0.5^{\phm{\star}}$ \\
    MACS J1311.0 & 0.49 & $\checkmark$ & & \multicolumn{4}{c}{no sources} \\
    MACS J1347.5 & 0.45 & $\checkmark$ & & 13:47:30.1 & $-$11:45:30.2 & 23 & $\phn17.7 \pm 3.2^{\phm{\star}}$ \\
                 &      &              & & 13:47:30.7 & $-$11:45:08.6 & $\phn2$ & $\phn45.9 \pm 1.5^{\phm{\star}}$ \\
    Abell 1835 & 0.25 & $\checkmark$ & & 14:01:02.1 & +02:52:41.0 & $\phn2$ & $\phn39.3 \pm 1.6^{\phm{\star}}$ \\
    MACS J1423.8 & 0.55 & $\checkmark$ & & 14:23:47.9 & +24:04:39.9 & $\phn3$ & $\phn\phn8.0 \pm 1.1^{\phm{\star}}$ \\
    MACS J1532.9 & 0.36 & $\checkmark$ & & 15:32:53.8 & +30:20:59.8 & $\phn1$ & $\phn22.8 \pm 0.8^{\phm{\star}}$ \\
    Abell 2204 & 0.15 & $\checkmark$ & & 16:32:46.9 & +05:34:34.9 & $\phn5$ & $\phn69.3 \pm 2.5^{\phm{\star}}$ \\
    Abell 2219 & 0.23 & & & 16:40:15.0 & +46:42:28.7 & 55 & $\phn\phn6.1 \pm 0.5^{\phm{\star}}$ \\
               &      & & & 16:40:21.8 & +46:42:47.8 & 24 & $239.1 \pm 8.3^{\phm{\star}}$ \\
               &      & & & 16:40:23.8 & +46:41:47.3 & 56 & $\phn\phn7.9 \pm 1.0^{\phm{\star}}$ \\
    MACS J1720.3 & 0.39 & $\checkmark$ & & 17:20:15.3 & +35:35:26.3 & 59 & $\phn\phn9.6 \pm 0.5^{\phm{\star}}$ \\
                 &      &              & & 17:20:16.8 & +35:36:28.4 & $\phn6$ & $\phn18.0 \pm 1.0^{\phm{\star}}$ \\
    Abell 2261 & 0.22 & $\checkmark$ & & 17:22:27.7 & +32:07:57.8 & $\phn8$ & $\phn\phn5.3 \pm 0.5^{\phm{\star}}$ \\
    MACS J1931.8 & 0.35 & $\checkmark$ & & 19:31:49.9 & $-$26:35:13.4 & 40 & $216.5 \pm 6.9^{\phm{\star}}$ \\
    MS 2053 & 0.58 & & $\checkmark$ & \multicolumn{4}{c}{no sources} \\
    MACS J2129.4 & 0.59 & & $\checkmark$ & \multicolumn{4}{c}{no sources} \\
    RX J2129.6 & 0.24 & $\checkmark$ & & 21:29:40.0 & +00:05:23.0 & $\phn7$ & $\phn25.4 \pm 1.2^{\phm{\star}}$ \\
    MS 2137 & 0.31 & $\checkmark$ & & 21:40:15.0 & $-$23:39:39.5 & $\phn2$ & $\phn\phn3.8 \pm 0.5^{\phm{\star}}$ \\
    MACS J2211.7 & 0.40 & $\checkmark$ & & \multicolumn{4}{c}{no sources} \\
    MACS J2214.9 & 0.50 & & $\checkmark$ & 22:14:56.5 & $-$14:00:55.7 & 46 & $\phn\phn5.6 \pm 0.6^{\phm{\star}}$ \\
    Abell S1063$^a$ & 0.35 & & & \multicolumn{4}{c}{no sources} 
   \enddata
   \tablecomments{NVSS sources within 1~arcmin of the cluster center
for the 45 clusters in our sample. The columns give the 
cluster name, the cluster redshift, the X-ray morphology,
the RA/dec of the NVSS source, the distance from the cluster
center to the NVSS source, and the NVSS
flux density at 1.4~GHz.
The superscript $a$ denotes that Abell S1063 is below the declination
range of the NVSS survey ($-40$ deg), and we therefore
searched the SUMSS catalog instead \citep{bock99, mauch03, murphy07}. 
The SUMSS survey is similar in depth
to the NVSS survey, although at lower frequency. We found no SUMSS sources within
1 arcmin of the center of Abell S1063.}
   \label{tab:radio_1}
\end{deluxetable*}

\begin{deluxetable*}{ccccccc} 
  \tabletypesize{\scriptsize}
  \tablewidth{\textwidth}
   \tablecaption{Central Radio Sources}
   \tablehead{\colhead{cluster} & 
     \colhead{NVSS} & \colhead{OVRO/BIMA} &
     \colhead{SZA} & \colhead{$\alpha_{1.4/30}$} &
     \colhead{extrap.} & \colhead{Bolocam} \\
     \colhead{} & \colhead{1.4 GHz} &
     \colhead{28.5 GHz} & \colhead{30.9 GHz} & \colhead{} &
     \colhead{140 GHz} & \colhead{140 GHz} \\
     \colhead{} & \colhead{mJy} &
     \colhead{mJy} & \colhead{mJy} & \colhead{} &
     \colhead{mJy} & \colhead{mJy}}
  \startdata
    MACS J0025.4 & $\phn28.7 \pm 1.3$ & $\phn0.33 \pm 0.17^c$ & & 
      $-1.38^{\phm{a}}$ & $0.06 \pm 0.05^{\phm{d}}$ & $-2.3 \pm 2.0$ \\
    ZWCL 0024 & $\phn\phn2.6 \pm 0.4$ & & $ \phm{} -0.2 \pm 0.2^{\phm{a}}$ & 
      $-1.09^{\phm{a}}$ & $0.01 \pm 0.02^{\phm{d}}$ & $\phm{-}0.6 \pm 5.3$ \\
    Abell 209 & $\phn18.4 \pm 1.0$ & & $\phn \phm{-} 1.1 \pm 0.2^{\phm{a}}$ & $-0.91^{\phm{a}}$ &
      $0.28 \pm 0.07^{\phm{d}}$ & $-1.2 \pm 2.1$ \\
    Abell 370 & $\phn10.5 \pm 1.0$ & $\phn0.72 \pm 0.09^c$ & $\phm{-}\phn0.8 \pm 0.2^{\phm{a}}$ & 
      $-0.88^{\phm{a}}$ & $0.18 \pm 0.03^{\phm{d}}$ & $-3.4 \pm 1.8$ \\
    Abell 383 & $\phn40.9 \pm 1.3$ & $\phn4.40 \pm 0.50^a$ & $\phm{-}\phn4.3 \pm 0.3^b$ & 
      $-0.73^{\phm{a}}$ & $1.41 \pm 0.14^{\phm{d}}$ & $\phm{-}5.4 \pm 2.1$ \\
    MACS J0329.6 & $\phn\phn6.9 \pm 0.6$ & & $\phm{-}\phn0.3 \pm 0.4^{\phm{a}}$ & 
      $-1.04^d$ & $0.10 \pm 0.12^{\phm{d}}$ & $\phm{-}3.9 \pm 3.1$ \\
    MACS J0417.5 & $\phn29.8 \pm 1.7$ & & & 
      $-1.11^d$ & $0.18 \pm 0.03^{\phm{d}}$ & $\phm{-}2.3 \pm 2.1$ \\
    MACS J0429.6 & $138.8 \pm 4.2$ & & $\phm{-}18.2 \pm 0.9^b$ & 
      $-0.66^{\phm{a}}$ & $6.70 \pm 0.50^{\phm{d}}$ & $\phm{-}9.8 \pm 3.5$ \\
    MS 1054 & $\phn14.1 \pm 0.9$ & $\phn0.94 \pm 0.07^a$ & & 
      $-0.90^{\phm{a}}$ & $0.22 \pm 0.03^{\phm{d}}$ & $\phm{-}1.6 \pm 1.7$ \\
    MACS J1115.8 & $\phn16.2 \pm 1.0$ & & $\phn \phm{-}1.4 \pm 0.4^b$ & 
      $-0.81^{\phm{a}}$ & $0.42 \pm 0.18^{\phm{d}}$ & $-2.7 \pm 1.9$ \\
    Abell 1423 & $\phn33.3 \pm 0.9$ & & $\phn \phm{-}0.6 \pm 0.2^{\phm{a}}$ & 
      $-1.31^{\phm{d}}$ & $0.09 \pm 0.04^{\phm{d}}$ & $-3.5 \pm 2.8$ \\
    MACS J1206.2 & $160.9 \pm 6.3$ & & & 
      $-1.38^d$ & $0.27 \pm 0.08^{\phm{d}}$ & $-5.4 \pm 1.9$ \\
    CL J1226 & $\phn\phn4.3 \pm 0.5$ & & $\phm{-}\phn0.3 \pm 0.2^{\phm{a}}$ & 
      $-0.92^{\phm{a}}$ & $0.09 \pm 0.08^{\phm{d}}$ & $\phm{-}1.3 \pm 3.3$ \\
    MACS J1347.5 & $\phn45.9 \pm 1.5$ & $10.38 \pm 0.47^a$ & $\phm{-}\phn8.7 \pm 0.5^b$ & 
      $-0.51^{\phm{a}}$ & $4.39 \pm 0.21^e$ & $-4.4 \pm 1.8$ \\
    Abell 1835 & $\phn39.3 \pm 1.6$ & $\phn2.97 \pm 0.13^a$ & $\phm{-}\phn2.9 \pm 0.3^b$ & 
      $-0.85^{\phm{a}}$ & $0.77 \pm 0.05^{\phm{d}}$ & $-1.1 \pm 1.4$ \\
    MACS J1423.8 & $\phn\phn8.0 \pm 1.1$ & $\phn1.49 \pm 0.13^a$ & $\phm{-}\phn2.0 \pm 0.2^b$ & 
      $-0.50^{\phm{a}}$ & $0.76 \pm 0.10^{\phm{d}}$ & $\phm{-}2.1 \pm 2.8$ \\
    MACS J1532.9 & $\phn22.8 \pm 0.8$ & $\phn3.25 \pm 0.22^a$ & $\phm{-}\phn3.2 \pm 0.3^b$ & 
      $-0.64^{\phm{a}}$ & $1.19 \pm 0.10^{\phm{d}}$ & $\phm{-}2.5 \pm 2.7$ \\
    Abell 2204 & $\phn69.3 \pm 2.5$ & $\phn8.79 \pm 0.37^a$ & $\phm{-}\phn7.0 \pm 0.4^b$ & 
      $-0.71^{\phm{a}}$ & $2.65 \pm 0.16^{\phm{d}}$ & $\phm{-}1.1 \pm 1.4$ \\
    Abell 2219 & $\phn\phn6.1 \pm 0.5$ & & $\phm{-}\phn0.6 \pm 0.2^{\phm{a}}$ & 
      $-0.74^{\phm{a}}$ & $0.21 \pm 0.09^{\phm{d}}$ & $\phm{-}4.9 \pm 2.6$ \\
     & $239.1 \pm 8.3$ & $14.87 \pm 0.62^a$ & $\phm{-}14.2 \pm 0.7^{\phm{a}}$ & 
      $-0.92^{\phm{a}}$ & $3.43 \pm 0.17^{\phm{d}}$ & $-4.9 \pm 3.3$ \\
     & $\phn\phn7.9 \pm 1.0$ & & $\phm{-}\phn0.1 \pm 0.1^{\phm{a}}$ & 
      $-1.40^{\phm{a}}$ & $0.02 \pm 0.02^{\phm{d}}$ & $-7.3 \pm 2.4$ \\
    MACS J1720.3 & $\phn\phn9.6 \pm 0.5$ & & $\phm{-}\phn0.6 \pm 0.1^{\phm{a}}$ & 
      $-0.90^{\phm{a}}$ & $0.15 \pm 0.04^{\phm{d}}$ & $\phm{-}5.3 \pm 1.6$ \\
     & $\phn18.0 \pm 1.0$ & & $\phm{-}\phn1.8 \pm 0.4^b$ & 
      $-0.76^{\phm{a}}$ & $0.58 \pm 0.21^{\phm{d}}$ & $-5.1 \pm 2.4$ \\
    Abell 2261 & $\phn\phn5.3 \pm 0.5$ & $\phn0.20 \pm 0.30^{\phm{d}}$ & & 
      $-1.10^{\phm{a}}$ & $0.05 \pm 0.07^{\phm{d}}$ & $-4.8 \pm 1.3$ \\
    MACS J1931.8 & $216.5 \pm 6.9$ & & & 
      $-0.72^d$ & $8.81 \pm 4.56^{\phm{d}}$ & $-3.0 \pm 2.2$ \\
    RX J2129.6 & $\phn25.4 \pm 1.2$ & $\phn2.33 \pm 0.14^a$ & $\phm{-}\phn2.6 \pm 0.2^b$ & 
      $-0.77^{\phm{a}}$ & $0.71 \pm 0.06^{\phm{d}}$ & $\phm{-}0.1 \pm 2.9$ \\
    MS 2137 & $\phn\phn3.8 \pm 0.5$ & & & 
      N/A & $0.06^f$  & $-7.5 \pm 5.1$ \\
    MACS J2214.9 & $\phn\phn5.6 \pm 0.6$ & & $\phm{-}\phn0.4 \pm 0.3^{\phm{a}}$ & 
      $-0.89^{\phm{a}}$ & $0.14 \pm 0.13^{\phm{d}}$ & $-2.8 \pm 2.3$ 
   \enddata
   \tablecomments{Compact central radio sources in our sample of 45 clusters.
     The columns give the
cluster name, the NVSS
flux density at 1.4~GHz, the OVRO/BIMA flux density at 28.5~GHz,
the SZA flux density at 30.9~GHz, the spectral index
determined from the 1.4 and $\simeq 30$~GHz data,
the predicted flux density at 140~GHz based on that 
spectral index,
and the measured flux density in our 140~GHz image
at the location of the source, which is in
some cases negative due to noise fluctuations.
All of the uncertainties are $1\sigma$ values and include
flux calibration uncertainties;
the Bolocam uncertainties are approximately Gaussian
and include degeneracies between the point source
and the cluster SZ signal (see text). 
Superscripts denote the following:
$a)$ measurements from \citet{coble07}, 
$b)$ measurements from \citet{bonamente11}, 
$c)$ sources published in \citet{coble07}, but refit 
in our analysis (the source associated with MACS J0025.4 was published
in B1950 coordinates in \citet{coble07}),
$d)$ spectral index based on fits with 4.85~GHz and/or
lower frequency data \citep{cohen07, large81, griffith94},
$e)$ source detected by two groups at mm/submm frequencies,
and the best-fit 140~GHz flux density from the combination
of those data and the radio results is $4.4 \pm 0.3$~mJy \citep{pointecouteau01}
and $5.5 \pm 0.6$~mJy \citep{komatsu99},
consistent with our extrapolated estimate,
$f)$ only 1.4~GHz data was available for this source, so it was extrapolated
based on the median value of $\alpha_{1.4/30}$ for our sample (-0.89).
We note that some of the extrapolated flux densities are consistent 
with 0. We quote symmetric uncertainties about these
best fit values, even though the true probability density
functions are asymmetric and exclude unphysical negative values.}
   \label{tab:radio_1b}
\end{deluxetable*}

\subsection{Central Radio Galaxies - 140 GHz Properties}

For each of these NVSS selected sources we have measured 
the 140~GHz flux density as follows.
First, we jointly fit a point source template centered on the
NVSS coordinates, 
along with a template
for the SZ emission, to our data.
Our fitting procedure is described in detail in \citet{sayers11},
and we refer the reader to that manuscript for additional details.
The normalization of the point-source template gives the
flux density of the source (in mJy).
For the SZ template, we used the morphology-dependent 
best-fit gNFW pressure profiles from \citet{arnaud10} (e.g., 
we used the best-fit cool-core
template for the cool-core clusters in our sample), allowing
the normalization of the template (i.e., $P_0$) to be a 
free parameter.
These profiles were used because all of the point sources 
we fit are in the central
regions of the cluster, where the shape of the pressure profiles 
varies
significantly with morphological type.
We then inserted the best-fit SZ and point-source templates
into 100 noise realizations and repeated the fit in order 
to characterize our uncertainty on the point-source
flux density. 
For most of our data the raw point-source sensitivity is
$\simeq 1$~mJy, but our uncertainties on the flux densities 
of these point sources are $\simeq 1-3$~mJy, with the 
degradation due to degeneracies between the SZ and point
source templates.
The sensitivity degradation depends strongly on the separation
of the SZ and point-source centroids, along with the 
angular size of the SZ template.
For clusters with multiple radio sources, we fit for
each source separately.
The best-fit flux densities are given in Table~\ref{tab:radio_1b}.

\begin{figure}
  \includegraphics[width=\columnwidth]{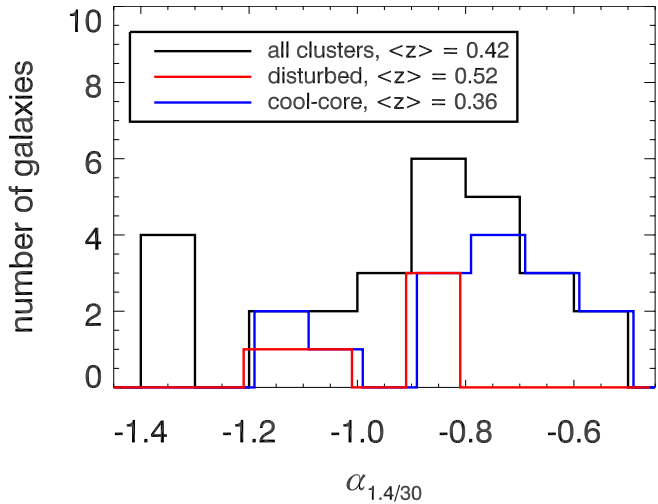}
  \caption{The value of the spectral index $\alpha_{1.4/30}$ found
    for the cluster-member galaxies in our sample using measurements
    at $\nu \le 30$~GHz. All of these
    galaxies are steep spectrum sources, and we find a median
    spectral index of $-0.89$.
    The spectral indices for the disturbed and cool-core
    sub-samples are also shown, and have been offset by 0.01 for 
    ease of viewing.
    Compared to the full sample, 
    the galaxies associated with cool-core clusters
    have shallower spectral indices on average.}
  \label{fig:alpha}
\end{figure}

We then derived an independent estimate for the 140~GHz flux densities
of the sources by extrapolating the low frequency data.
We have primarily combined 1.4~GHz NVSS data
with $\simeq 30$~GHz
measurements from SZA and/or OVRO/BIMA, which
exist for 24/28 sources \citep{bonamente11, coble07}.\footnote{
3/31 of the NVSS sources are extended and appear to have very
steep spectral indexes. Extrapolating to 140~GHz, we expect
negligible flux density from these three sources and therefore
have not attempted to fit to them in our data.}
Three of the sources without 30~GHz data have measurements at either
4.85~GHz and/or frequencies below 1.4~GHz, which have been
used to constrain the spectral energy distributions (SEDs) of 
those sources \citep{cohen07, large81, griffith94}.
We fit an SED of the form $\nu^{\alpha_{1.4/30}}$, and we find all of the
galaxies have $\alpha_{1.4/30} < -0.50$ (see Figure~\ref{fig:alpha}).
Our fits include both the measurement uncertainties and overall
flux calibration uncertainties on the source flux densities,
the latter of which is equalt to 3\% for the NVSS measurements, 
4\% for the OVRO/BIMA measurements, and 5\% for SZA measurements 
\citep{condon98, reese02, muchovej07}.
We note that there are some systematic differences
in calibration between the datasets (e.g., using pre- or post-WMAP
planetary models), but these differences are below the
flux calibration uncertainties quoted for each dataset.
Furthermore, for our SED fits we assume a single observing frequency for each
dataset independent of source spectrum. This results in a negligible bias in
our results due to the small fractional bandwidths and
large separations in band centers among our datasets (1.4, 30, and 140~GHz).

Although our source selection using 1.4~GHz data favors steep spectrum
sources, we note that the $\simeq 30$~GHz data do not contain any
point source detections without an NVSS counterpart.
The $\simeq 30$~GHz data are generally sensitive to sources
with flux densities larger than $0.5 - 1.0$~mJy, and would therefore have detected
bright flat spectrum sources if they exist in the central regions
of these clusters.
Since all of the bright central radio sources are steep spectrum sources, which
tend to have little or no temporal variability \citep{tingay03, sadler06, bolton06}, 
the asynchronous
NVSS/SZA/OVRO/BIMA observations should provide an accurate
characterization of the source SED.
We used the values of $\alpha_{1.4/30}$ to predict the flux density at 140~GHz,
and these extrapolated values are given in Table~\ref{tab:radio_1b}.

None of the NVSS sources near cluster centers are detected at
a significant level in our 140~GHz images ($> 3\sigma$).
Consequently, we have characterized the average point-source signal
measured in our data by comparing our measured flux density
to the predicted flux density based on the extrapolation
from lower frequency measurements.
To avoid correlations and biases in this average measurement,
we have discarded the data from clusters with multiple NVSS
sources, those with no 30~GHz data, and also Abell 2261 due
to the fact that the SZ template does not provide a good
description of those data.
For the 17 remaining clusters we formed the quantity
\begin{displaymath}
\delta = \frac{S_{\textrm{Bolo}} - S_{\textrm{extrap}}}
  {\sqrt{\sigma^2(S_{\textrm{Bolo}}) + \sigma^2(S_{\textrm{extrap}})}}, 
\end{displaymath}
where $S_{\textrm{Bolo}}$ is the best-fit flux density in our 
Bolocam image, $S_{\textrm{extrap}}$ is the extrapolated flux
density based on the lower frequency measurements assuming
a constant spectral index, $\sigma(S_{\textrm{Bolo}})$ is the
uncertainty on the Bolocam measurement, and 
$\sigma(S_{\textrm{extrap}})$ is the uncertainty on the 
extrapolated flux density.

If our extrapolation of the spectral index found at lower
frequencies perfectly describes our data then the values
of $\delta$
will have a mean of 0 and standard deviation of 1.
We find a mean
$\delta = -0.26 \pm 0.28$, 
which indicates that 
our data show no statistically significant evidence for a change
in the spectral index above 30~GHz,
although our data are consistent with a slight steepening
of the spectral index
(a steepening is consistent with previous results and theoretical
expectations \citep{tucci11, ricci06, kellermann66}).
Furthermore, we note that a one-sample Kolmogorov-Smirnov goodness of fit
test shows
that our values of $\delta$ are consistent with 
a Gaussian distribution with a probability to exceed of 0.60.
In addition, we directly computed the standard deviation of the 17 values
of $\delta$, and we find that this standard deviation is 
$\textrm{StdDev}({\delta}) = 1.14 \pm 0.18$. 
This value is consistent with 1, which is the expectation under
the assumption that the dispersion in the values of $\delta$ is
due entirely to measurement noise.
However, if we assume there is an intrinsic scatter $\sigma_{\textrm{int}}$
in the true 140~GHz flux densities compared to the extrapolated
140~GHz flux densities, then
\begin{displaymath}
  \textrm{StdDev}({\delta}) = \sqrt{1 + \sigma_{\textrm{int}}^2},
\end{displaymath}
and our data imply a best-fit intrinsic scatter of $\simeq 30$\%.

\begin{deluxetable*}{ccccccc} 
  \tabletypesize{\scriptsize}
  \tablewidth{\textwidth}
   \tablecaption{Compact Central Radio Sources}
   \tablehead{\colhead{cluster} & 
     \colhead{subtracted PS} & \colhead{SZ S/N (raw)} & \colhead{$\theta_c$ (raw)} &
     \colhead{SZ S/N (PS-sub)} & \colhead{$\theta_c$ (PS-sub)} & \colhead{$\Delta$ S/N } \\
     \colhead{} & \colhead{mJy} & \colhead{} & \colhead{arcmin} & \colhead{} & \colhead{arcmin} & \colhead{percent}}
   \startdata
    MACS J0025.4 & $0.06^{\phm{a}}$ & 12.31 & 0.50 & 12.33 & 0.50 & $\phn$0.2 \\
    ZWCL 0024 & $0.01^{\phm{a}}$ & $\phn$3.26 & 3.00 & $\phn$3.26 & 3.00 & $\phn$0.0 \\
    Abell 209 & $0.28^{\phm{a}}$ & 13.93 & 1.75 & 14.00 & 1.50 & $\phn$0.5 \\
    Abell 370 & $0.18^{\phm{a}}$ & 12.80 & 0.50 & 12.89 & 0.50 & $\phn$0.7 \\
    Abell 383$^a$ & $1.41^{\phm{a}}$ & $\phn$9.33 & 2.00 & $\phn$9.72 & 2.00 & $\phn$4.2 \\
    MACS J0329.6 & $0.10^{\phm{a}}$ & 12.09 & 0.25 & 12.13 & 0.25 & $\phn$0.3 \\
    MACS J0417.5 & $0.18^{\phm{a}}$ & 22.67 & 1.00 & 22.71 & 1.00 & $\phn$0.2 \\
    MACS J0429.6 & $6.70^{\phm{a}}$ & $\phn$7.34 & 1.50 & $\phn$9.09 & 1.00 & 23.8 \\
    MS 1054 & $0.22^{\phm{a}}$ & 17.38 & 0.50 & 17.55 & 0.50 & $\phn$1.0 \\
    MACS J1115.8 & $0.42^{\phm{a}}$ & 10.92 & 1.25 & 11.05 & 1.25 & $\phn$1.2 \\
    Abell 1423 & $0.09^{\phm{a}}$ & $\phn5.80$ & 0.50 & $\phn5.82$ & 0.50 & $\phn0.3$ \\
    MACS J1206.2 & $0.27^{\phm{a}}$ & 21.69 & 1.00 & 21.80 & 1.00 & $\phn$0.5 \\
    CL J1226 & $0.09^{\phm{a}}$ & 13.01 & 0.25 & 13.05 & 0.25 & $\phn$0.3 \\
    MACS J1347.5 & $4.39^{\phm{a}}$ & 34.04 & 0.50 & 36.90 & 0.25 & $\phn$8.4 \\
    Abell 1835 & $0.77^{\phm{a}}$ & 15.34 & 0.75 & 15.76 & 0.50 & $\phn$2.8 \\
    MACS J1423.8 & $0.76^{\phm{a}}$ & $\phn$8.98 & 0.50 & $\phn$9.39 & 0.50 & $\phn$4.6 \\
    MACS J1532.9 & $1.19^{\phm{a}}$ & $\phn$7.55 & 0.75 & $\phn$8.04 & 0.50 & $\phn$6.5 \\
    Abell 2204 & $2.65^{\phm{a}}$ & 20.94 & 1.00 & 22.41 & 0.75 & $\phn$7.0 \\
    Abell 2219 & $3.43^{\phm{a}}$ & 10.53 & 1.00 & 11.12 & 1.00 & $\phn$5.6 \\
     & $0.21^{\phm{a}}$ & & & & & \\
     & $0.02^{\phm{a}}$ & & & & & \\
    MACS J1720.3 & $0.58^{\phm{a}}$ & 10.32 & 0.25 & 10.67 & 0.25 & $\phn$3.4 \\
     & $0.15^{\phm{a}}$ & & & & & \\
    Abell 2261$^a$ & $0.05^{\phm{a}}$ & 10.76 & 0.25 & 10.79 & 0.25 & $\phn$0.3 \\
    MACS J1931.8 & $8.81^{\phm{a}}$ & $\phn$9.21 & 1.25 & 10.20 & 1.00 & 10.8 \\
    RX J2129.6 & $0.71^{\phm{a}}$ & $\phn$7.76 & 0.75 & $\phn$8.02 & 0.75 & $\phn$3.4\\
    MS 2137 & $0.06^{\phm{a}}$ & $\phn$6.51 & 0.25 & $\phn$6.53 & 0.25 & $\phn$0.3 \\
    MACS J2214.9$^a$ & $0.14^{\phm{a}}$ & 12.85 & 1.00 & 12.86 & 1.00 & $\phn$0.1 
\enddata
\tablecomments{Compact central radio sources subtracted from Bolocam cluster data
  based on extrapolations from lower frequencies.
  From left to right the columns give:
  the flux density of the subtracted radio source, the SZ S/N in
  our unsubtracted image, the filter scale corresponding to this
  peak SZ S/N in our unsubtracted image, the SZ S/N in our 
  point-source subtracted image, the filter scale corresponding
  to the peak SZ S/N in our point-source subtracted image,
  and the fractional difference in SZ S/N between the unsubtracted
  and point-source subtracted images.
  In general, the extrapolated flux densities of the radio sources
  are accurate to $\simeq 30$\%, limited by the intrinsic scatter
  in the extrapolation.
  Consequently, the fractional change in the SZ S/N is also
  accurate to $\simeq 30$\%.
  The superscript $a$ denotes clusters that contain point sources 
  directly detected in our Bolocam
  images away from the cluster centers
  (see Section~\ref{sec:five}). These additional point sources 
  were not subtracted for this analysis.}
\label{tab:radio_2}
\end{deluxetable*}

\begin{figure*}
  \includegraphics[height=.33\textwidth]{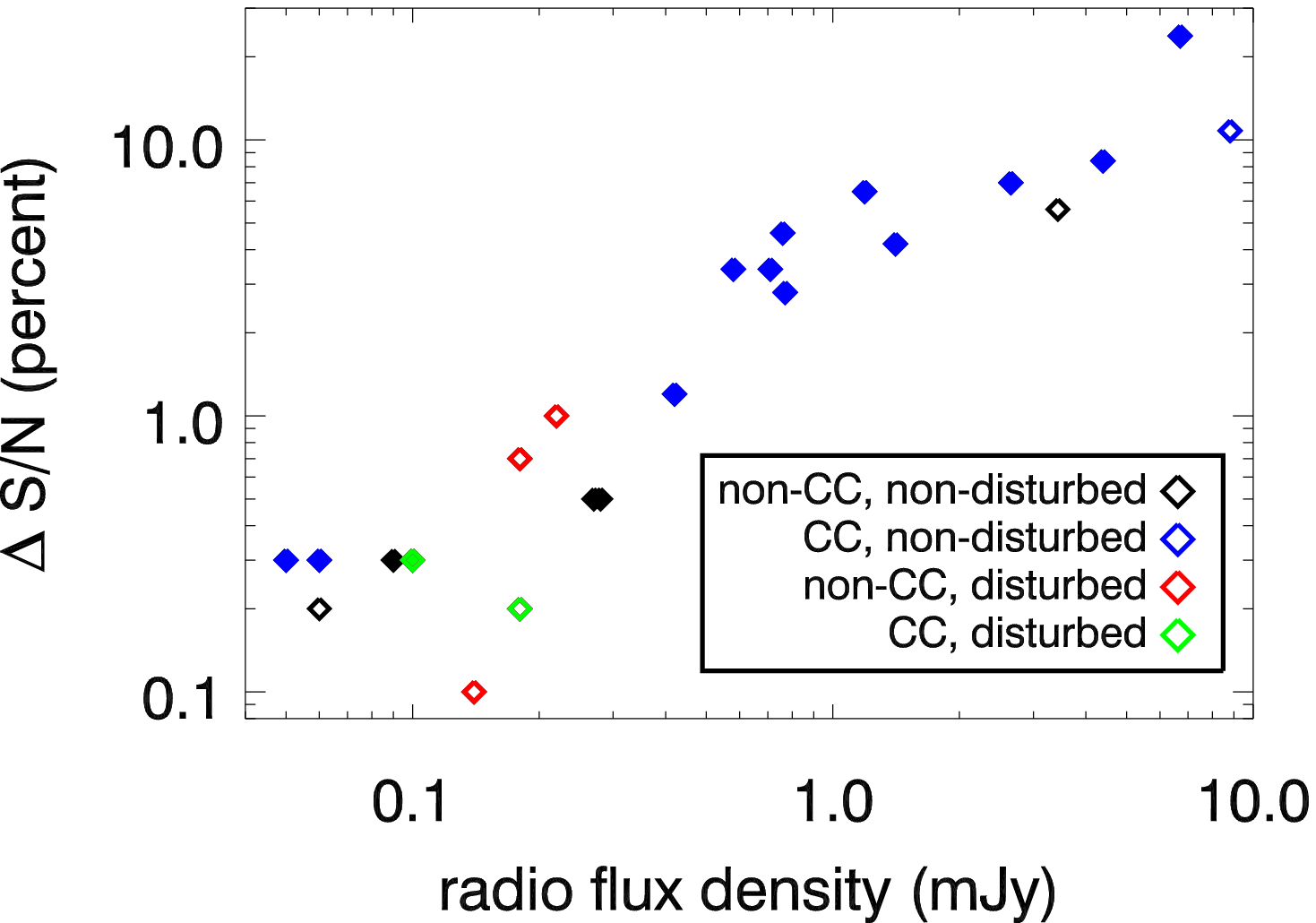}
\hspace{.05\textwidth}
  \includegraphics[height=.33\textwidth]{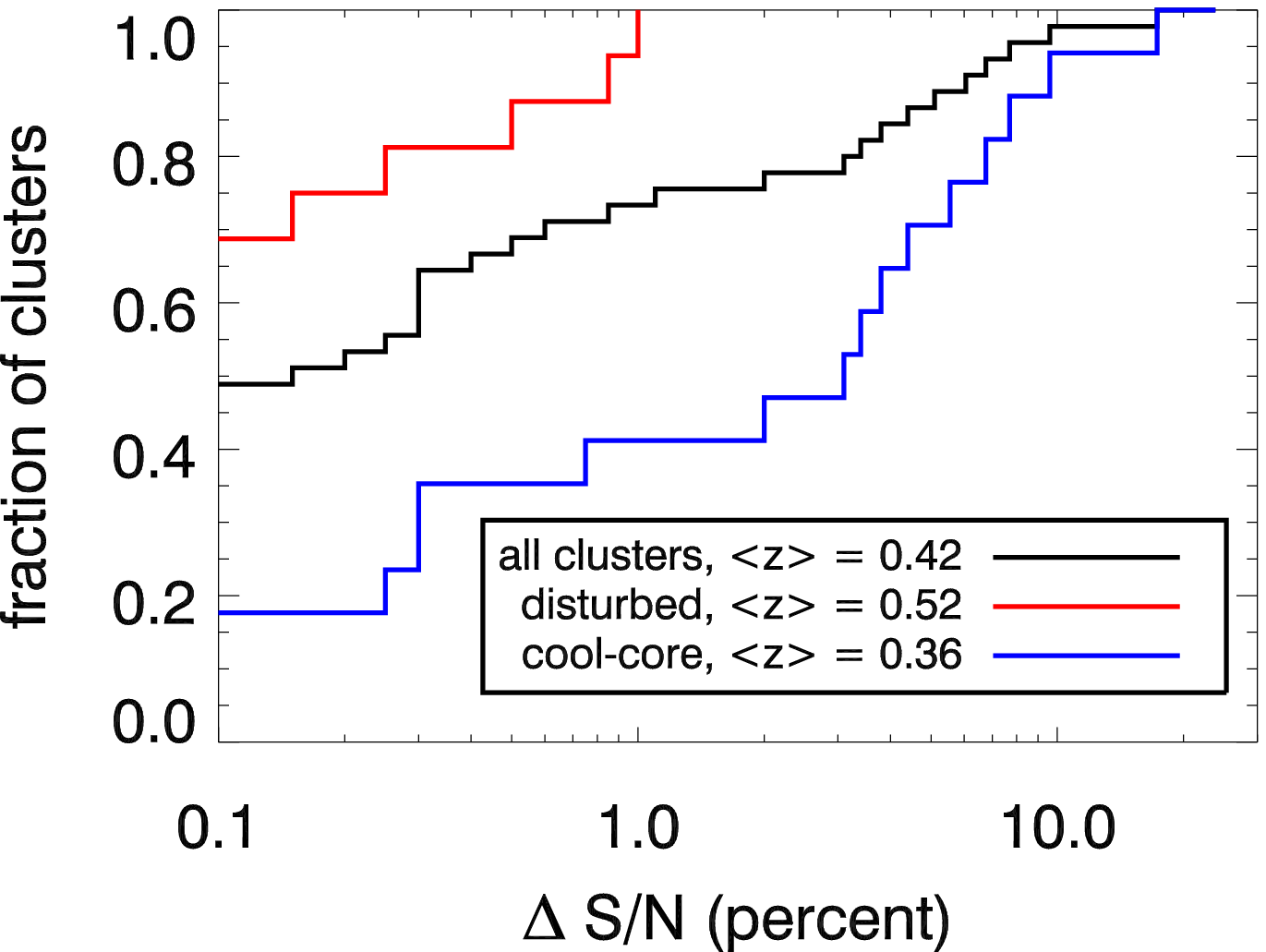}

\vspace{.05\textwidth}

  \includegraphics[height=.33\textwidth]{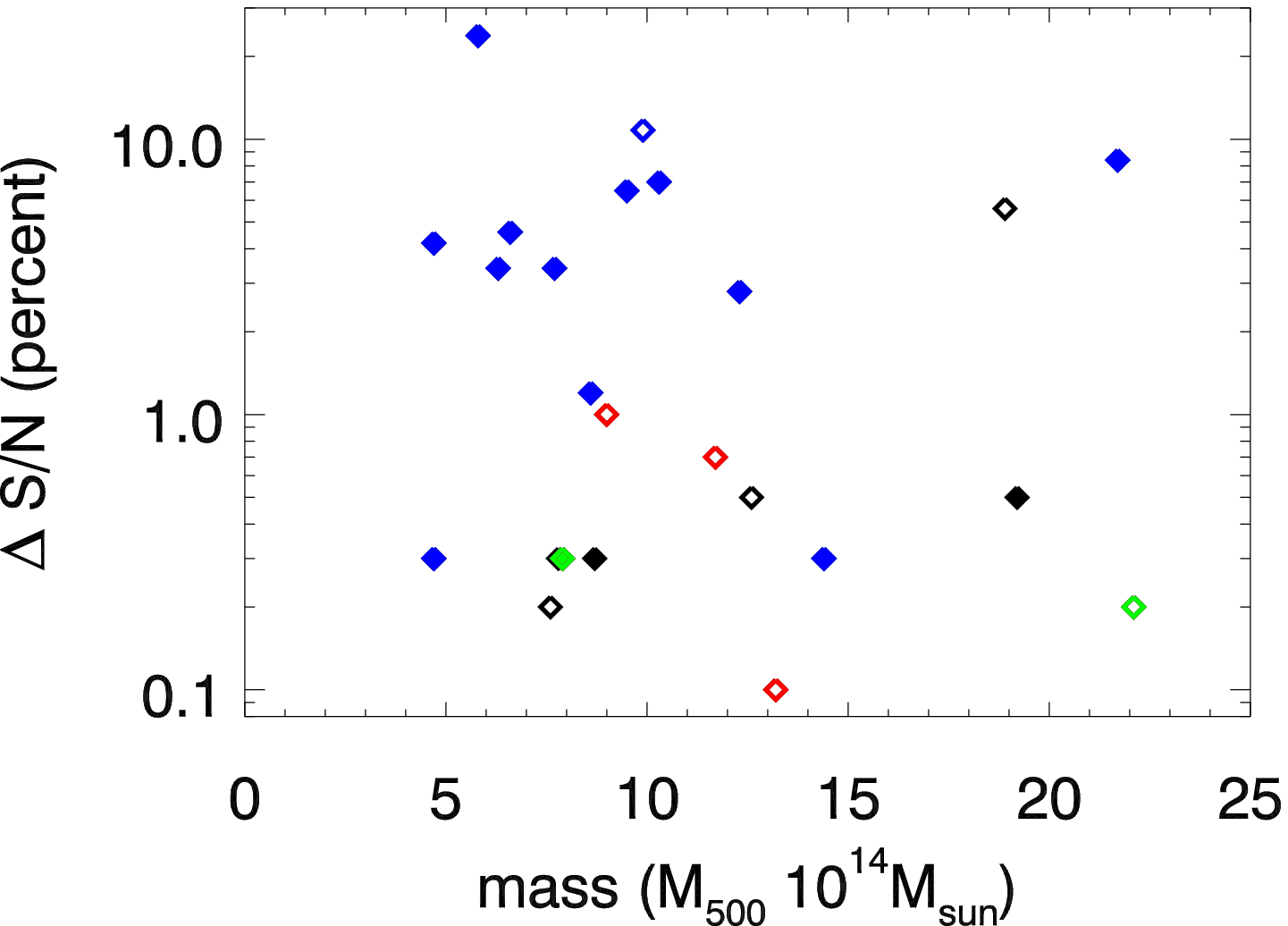}
\hspace{.06\textwidth}
  \includegraphics[height=.33\textwidth]{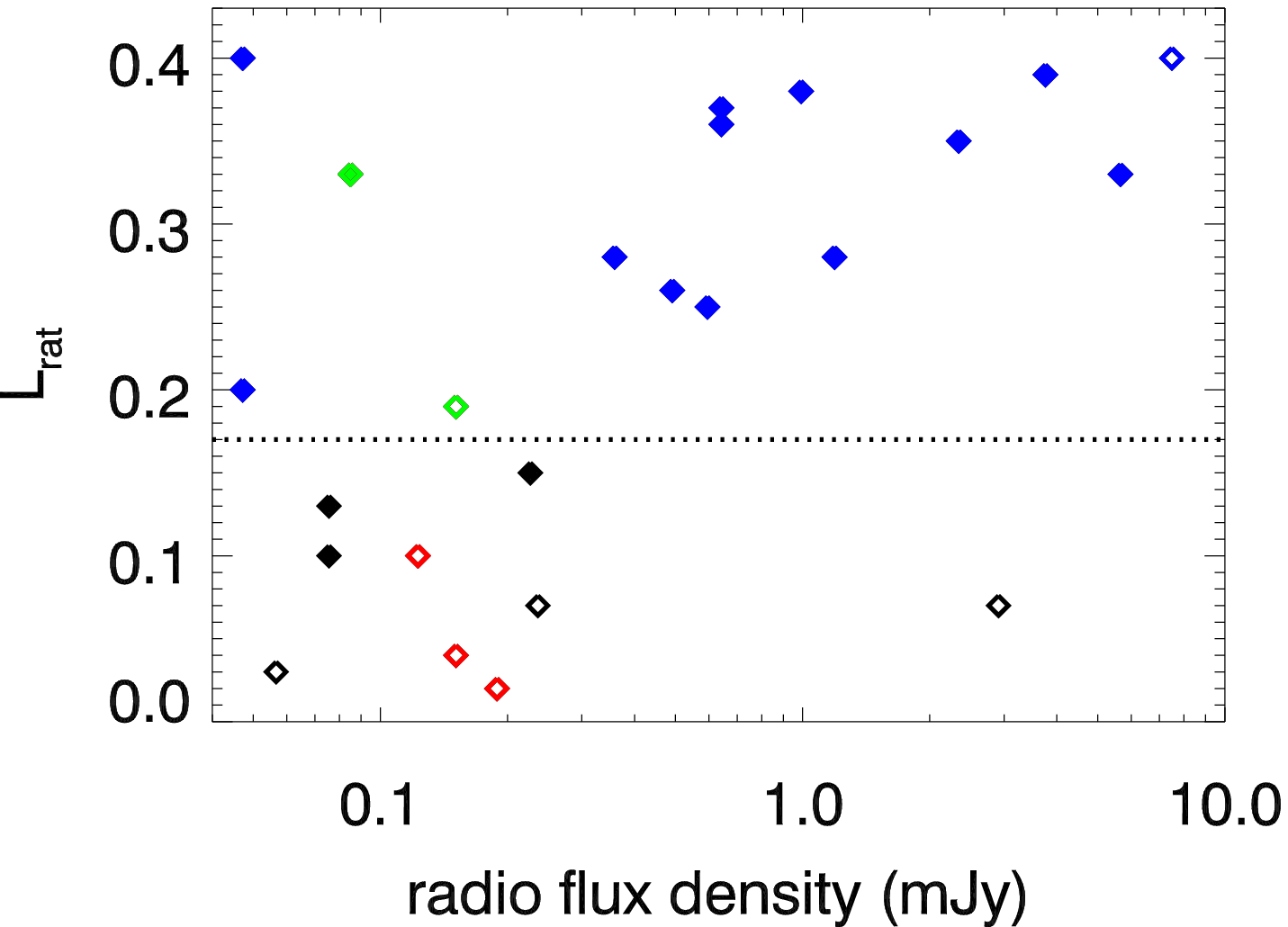}

  \caption{\scriptsize Top left: The fractional change in the SZ S/N
    as a function of the extrapolated 140~GHz flux density
    of the radio source. The maximum change is $\simeq 20$\%
    and a flux density $\gtrsim 0.5$~mJy is required to
    produce more than a 1\% change in SZ S/N.
    The data are color-coded by X-ray derived morphology,
    and nearly all of the brightest radio sources
    are associated with cool-core systems.
    Over half of the radio sources are within 10~arcsec of the
    cluster center ($25 - 75$~kpc for the redshift range of 
    our sample), and these sources are marked as filled symbols.
    Note that roughly half the clusters in our sample contain
    no radio sources given our selection criteria, and are
    therefore not included in this plot.
    Top right: The cumulative fraction of clusters with radio contamination below
    a given threshold (as quantified by fractional change
    in SZ S/N). 
    The black, red, and blue curves represent the full
    sample of 45 clusters, the disturbed sub-sample,
    and the cool-core subsample.
    The radio contamination is severe enough
    to cause a $> 1$\% change in the SZ signal for
    $\simeq 25$\% of the clusters in our sample,
    although the contamination is significantly worse
    in cool-core systems.
    Note that the cumulative fraction ignores sources
    undetected by NVSS, and is therefore likely to be
    an overestimate at $\Delta$ S/N $\lesssim 1$\%.
    Bottom left: The fractional change in SZ S/N versus
    cluster mass, showing no evidence for a correlation
    and indicating that the fractional amount of radio
    contamination is approximately independent of mass
    for these high mass clusters.
    Bottom right: The X-ray projected luminosity ratio
    $\textrm{L}_{\textrm{rat}} = \textrm{L}(\textrm{R} < 0.05\textrm{R}_{500}) / 
    \textrm{L}(\textrm{R} < \textrm{R}_{500})$ versus
    the 140~GHz flux density of the radio source, which
    shows a correlation and indicates that the 
    brightest radio sources are preferentially found
    in the clusters with the largest projected luminosity ratio.
    We define cool-core systems as clusters with $\textrm{L}_{\textrm{rat}} \ge 0.17$
    \citep{mantz09}.}
  \label{fig:one}
\end{figure*}

We also examined whether the Bolocam data show a significant detection
of the ensemble-average flux density of these radio sources,
again focusing on the same subset of 17 objects.
Analogous to the quantity $\delta$, we formed the quantity
\begin{displaymath}
  \textrm{SN}_{\textrm{Bolo}} = \frac{S_{\textrm{Bolo}}}{\sigma(S_{\textrm{Bolo}})}.
\end{displaymath}
We find a mean $\textrm{SN}_{\textrm{Bolo}} = 0.14 \pm 0.24$, indicating that
our 140~GHz data do not show statistically significant evidence for
the presence of these radio sources.
However, if we focus solely on the brightest sources with $S_{\textrm{extrap}} > 1$~mJy,
then we find a mean $\textrm{SN}_{\textrm{Bolo}} = 1.76 \pm 0.50$,
showing a detection of these sources at modest statistical significance ($\simeq 3.5\sigma$).
This implies that our non-detection of the ensemble-average of all radio
sources is likely due to the fact that most of the sources are well below
the Bolocam noise RMS, which means that averaging over these sources adds a significant
amount of noise but very little signal.
For completeness, we note that the mean value of $\delta$ for these brightest
sources is $0.54 \pm 0.62$, which is also consistent with
no change in the spectral index between 30~GHz and 140~GHz.

\subsection{Central Radio Galaxies - SZ Contamination}
\label{sec:central}

Since the extrapolation of the radio galaxy flux densities 
from lower frequencies provides a good description of our
140~GHz data, we subtracted each of the radio galaxies using
the extrapolated flux density.
A complete list of the subtracted radio sources is given in
Table~\ref{tab:radio_2}.
In most cases the measurement uncertainty on the extrapolated 
flux density is quite small, and our overall uncertainty
is therefore dominated by the 30\% intrinsic scatter in the
extrapolation.
We then computed the peak SZ S/N for each cluster both
with and without the radio source(s) subtracted using the 
optimal filtering formalism described in \citet{sayers12_3}.
Briefly, we construct an SZ template of the form
$S(\theta) \propto (1 + \theta^2/\theta_c^2)^{-1}$, weight
this template by the signal transfer function of our map
and the inverse of the map variance, convolve our map with
the resulting filter, and search for the peak S/N. 
This is repeated for $\theta_c = 0.25, 0.50,..,3.00$, and
the maximum peak S/N over all filter scales is used
to describe the SZ S/N.
This is essentially the same algorithm used by both ACT and 
SPT to search for clusters in their survey data 
\citep{vanderlinde10, marriage11a, reichardt12}.
We find the maximum fractional change in SZ S/N for any of the
clusters is $\simeq 20$\%,
and only 13/45 of the clusters show a fractional change in SZ S/N
larger than 1\% (see Figure~\ref{fig:one}).
Consequently, for the high-mass and moderate-redshift clusters
in our sample contamination from radio emission is in 
general small compared to the noise in our data.
Furthermore, for the brightest radio galaxies we generally find that the
value of $\theta_c$ with the maximum SZ S/N is smaller when
the radio source is subtracted from the map, which is due to the 
resulting cluster profile being more sharply peaked
(recall that these radio sources partially fill in the SZ
decrement at 140~GHz).

Although the overall level of radio contamination in our sample
is small, we note that 11/12 of the brightest radio sources
are associated with cool-core systems, and therefore
11/17 cool-core systems show a fractional change in SZ S/N
larger than 1\%.\footnote{
  The median redshift of the cool-core systems in our
  sample ($z = 0.36$) is slightly lower than the median
  redshift of the full sample ($z = 0.42$) and
  the median redshift of the disturbed systems ($z=0.52$).
  Consequently, some of the difference in radio
  contamination for the different morphological
  classifications may be due to these
  differences in redshift rather than the cluster
  environment.}
Previous studies have found a similar relationship
between strong radio emission and cool-core systems
(e.g., \citet{sun09}).
Further supporting this trend of cool-core systems hosting 
the brightest radio galaxies,
we also find a correlation between the strength
of the radio emission and the projected X-ray luminosity
ratio, which we have used to separate cool-core and 
non-cool-core systems (see Figure~\ref{fig:one}).
This correlation is quantified by a Spearman rank coefficient
of 0.46 and null hypothesis probability of 0.02 
(for 25 data points).
In addition, we find that there is no significant correlation
between the fractional change in SZ S/N and cluster mass for our sample,
quantified by a Spearman rank coefficient of -0.11 and a
null hypothesis probability of 0.59 (again for 25 data points,
see Figure~\ref{fig:one}).
Consequently, the fractional change in SZ S/N due to radio
contamination is approximately independent of mass for these
high mass systems.
Since there is no evidence for a correlation between cluster
mass and fractional change in SZ S/N, and since
our sample approximately spans the mass range of objects
discovered by the ACT and SPT surveys \citep{marriage11a, reichardt12}, 
the amount of radio contamination in those surveys should
be similar to what we have found for our sample.
However, we caution that those surveys contain a significantly
larger fraction of low-mass clusters, and the negative Spearman
rank coefficient indicates that our data favor
an on-average increase in fractional change in SZ S/N with
decreasing cluster mass (although at low significance
given the 59\% probability
of our data being consistent with no trend in mass).

Due to the way we have selected the cluster-member radio
sources, anything below the NVSS detection limit (2.5 mJy at 1.4 GHz)
will not be included in our analysis.
However, even if we assume the smallest magnitude spectral index
found for any of the radio galaxies in our sample ($\alpha_{1.4/30} = -0.50$),
a source undetected by NVSS would have a maximum 140~GHz flux density
of $0.25$~mJy.
Therefore, it is unlikely that a source undetected by NVSS
would have more than a 1\% effect on the optimally filtered
peak SZ signal we 
measure for any of the clusters in the Bolocam sample.
We do note that, of the 21 central radio sources detected by
\citet{coble07}, there were two sources with spectral indices 
smaller in magnitude than $\alpha_{1.4/30} = -0.5$, and one source
with a spectral index of $\alpha_{1.4/30} = -0.14$.
Since our $\simeq 30$~GHz data are most sensitive to these flat spectrum
sources, we consider the the extreme case of a source with $\alpha_{1.4/30} = -0.14$
and a $\simeq 30$~GHz flux density just below our detection limit of $\lesssim 1$~mJy.
Such a source could go undetected in our $\simeq 30$~GHz data and still 
have a 140~GHz flux density of $\simeq 0.5$~mJy.
Such a source would likely produce a 
$\simeq 1$\% change in the SZ S/N measured by Bolocam (see Figure~\ref{fig:one}).
However, given the rarity of such flat spectrum sources in the 
centers of clusters, coupled with the fact that the source
would also need to be close to our detection threshold,
it is unlikely that there are a significant number
of such sources in the Bolocam cluster sample.

\subsection{Non-Central Cluster-Member Radio Galaxies}

Although the number of cluster radio galaxies drops quickly with radius, there are 
likely to be some cluster member radio galaxies outside of our 1~arcmin
diameter search radius.
For example, the total number of radio sources in the central $\simeq 2$~arcmin radius
of our images is enhanced by a 
factor of $\simeq 3$ compared to the blank-sky average \citep{coble07, muchovej10}.
However, there are significantly fewer sources in these regions compared
to the cluster centers (a factor of $\simeq 10$),
and the contamination from these sources will be
much less degenerate with the cluster SZ signal due to the relatively
large spatial separation.
Therefore, we expect the contamination from these sources to be 
significantly smaller than
the contamination from the central radio galaxies.
Consequently, we examine the effects of these sources in a 
statistical sense.

First, the factor of 3 increase in the number of radio sources will increase
the corresponding noise fluctuations from undetected sources
by a factor of $\simeq \sqrt{3}$.
However, even with this enhancement the noise fluctuations from
radio sources will still be sub-dominant to the noise fluctuations
from dusty star-forming galaxies and negligible compared to
the noise RMS in our images ($\lesssim 0.1$~mJy compared to $\simeq 1$~mJy).
Furthermore, we note that the increased number of radio sources in the 
direction of the cluster
will systematically fill in the SZ decrement, which is referenced to the average 
signal level outside of the cluster, and bias our results.
Based on a simple power-law extrapolation of the measured number 
counts at $\simeq 140$~GHz,\footnote{
Measurements at a wide range of frequencies indicate that
a simple power law is a good approximation of the number counts
on the low-flux side of the break in the counts (e.g., \citet{tucci11}).} 
which constrain the number counts above $\simeq 10$~mJy,
we estimate that the SZ decrement
will be filled in by $\simeq 0.01$~mJy/arcmin$^2$ due to the 
higher than average number of radio sources towards the cluster
\citep{vieira10, marriage11, planck11_xiii}.
This surface brightness is almost 3 orders of magnitude below the average
SZ surface brightness towards the clusters in our sample and therefore
should produce a $\simeq 0.1$\% bias in the average SZ signal
recovered.
Consequently, contamination from radio galaxies outside of 1~arcmin in
radius from the cluster center should generally be negligible.

\section{Possible Contamination from Dusty Galaxies}

In addition to radio emission, some of the cluster-member galaxies
are likely to have thermal dust emission.
Studies of clusters out to $z=0.8$
in the mid-IR with \emph{Spitzer} have shown 
that there can be significant enhancements in the number
of 24 $\mu$m-detected sources within narrow redshift slices centered on
clusters \citep{bai07, marcillac07}.
However, the total number of 24 $\mu$m
objects detected towards clusters is $\lesssim 10$\%
above the number detected towards blank fields at similar
sensitivities, due to the fact that the
sources are detected over a wide range of redshifts \citep{geach06, finn10}.
Compared to 24~$\mu$m observations, our data are significantly more sensitive
to high-redshift galaxies and less sensitive to dust emission
from cluster-member galaxies \citep{blain02,bethermin11}, and so
the effective enhancement compared to a blank field will be
even smaller in our Bolocam images.
Furthermore, \emph{Herschel} has recently published results characterizing
the dust emission from 68 cluster BCGs at frequencies as low as
600~GHz \citep{rawle12}.
15 of the BCGs are detected in one or more Herschel bands, and
grey-body fits of the \emph{Herschel} data result
in extrapolated flux densities at our observing band of $<0.1$~mJy
for all but two of these sources.
Based on our analysis of the contamination from central radio galaxies,
even these brightest dusty sources would result in $< 1$\% fractional changes
in the SZ S/N.
In summary, the dust emission sourced by cluster-member galaxies
in our images is likely to be negligible, and we therefore
do not account for it in this analysis.

\section{Radio Sources Directly Detected by Bolocam}

\label{sec:five}

\begin{deluxetable*}{ccccccccc} 
  \tabletypesize{\tiny}
  \tablewidth{\textwidth}
   \tablecaption{Point Sources Detected by Bolocam}
   \tablehead{\colhead{cluster} & \colhead{coordinates} & \colhead{radius} &
     \colhead{NVSS} & \colhead{OVRO/BIMA} & \colhead{SZA} & \colhead{Bolocam} &
     \colhead{SZ S/N} & \colhead{SZ S/N} \\
     \colhead{} & \colhead{J2000} & \colhead{arcmin} & \colhead{1.4 GHz} & \colhead{28.5 GHz} &
     \colhead{30.9 GHz} & \colhead{140 GHz} & \colhead{(raw)} & \colhead{(PS-sub)}}
   \startdata
    Abell 267 & 01:52:54.6, +01:02:08.2 & 3.51 & $\phn\phn\phn4.6 \pm \phn0.5$ & $\phn\phn7.55 \pm 0.24$ & & $\phn9.8 \pm 2.3$ & \phn9.83 & \phn9.57 \\
    Abell 383 & 02:48:22.1, $-$03:34:30.5 & 5.44 & $\phn\phn54.9 \pm \phn2.4$ & & $7.5 \pm 0.3$ & $14.3 \pm 1.7$ & \phn9.56 & \phn9.57\\
    Abell 963 & 10:17:14.2, +39:01:24.0 & 2.53 & $1392.2 \pm 41.8$ & & & $26.8 \pm 2.8$ & \phn9.32 & \phn8.31 \\
    Abell 2261 & 17:22:16.9, +32:09:10.4 & 2.46 & $\phn\phn24.3 \pm \phn1.6$ & $\phn10.48 \pm 0.16$ & & $\phn8.7 \pm 1.1$ & 10.79 & 10.18 \\
               & 17:22:23.8, +32:01:26.4 & 6.56 & $\phn126.7 \pm \phn4.4$ & & & $\phn9.3 \pm 3.4$ & \\
    MACS J2214.9 & 22:14:39.3, $-$14:00:58.5 & 4.44 & $\phn\phn58.4 \pm \phn1.8$ & $107.10 \pm 0.58$ & & $51.4 \pm 5.2$ & 12.86 & 12.60
   \enddata
\tablecomments{Point sources detected in Bolocam cluster observations with S/N $>4$
  in SZ-model-subtracted beam-smoothed images.
  The uncertainties given above fully account for degeneracies between the point
  source and the SZ model, and are therefore larger than the noise RMS in the
  beam-smoothed images.
  As a result, our constraint on the flux density of one source has a S/N $<4$.
  Note that, in contrast to the central point sources 
  selected from NVSS, 5/6 of these sources appear to be flat-spectrum sources.
  This is due to the fact that most steep-spectrum sources are too dim
  to be detected in our 140~GHz data.}
\label{tab:radio_3}
\end{deluxetable*}

The number of cluster member radio galaxies drops quickly
with radius from the cluster center \citep{coble07, muchovej10},
and the blank sky description of the source population
becomes approximately correct at these larger radii.
Over the full set of 45 clusters we map a fairly significant
area ($\simeq 2.5$~deg$^2$), which is enough
to contain a handful of very bright radio sources \citep{vieira10, marriage11}.
Fortunately, we are able to easily detect and subtract these sources from
our data because they have little or no degeneracy with our template of the
SZ signal.
We search for these sources by selecting pixel excursions with a S/N $>4$
in maps that have had the best-fit SZ template subtracted and
are beam-smoothed for improved point source extraction.
We find 15 such excursions, 12 with positive flux density and
three with negative flux density.
Six of these positive flux candidates
have counterparts in NVSS, and we estimate the 140~GHz flux
density of each of these sources by fitting a point source
template to our maps using the NVSS coordinates
as a prior \citep{condon98}.
The dimmest of the sources with NVSS associations have
flux densities of 
$\lesssim 10$~mJy in our images, providing a rough estimate 
of our detection threshold (see Table~\ref{tab:radio_3}).
Although our detection threshold is not uniform due to
residual SZ contamination, the varying depths of our maps,
and the tapered coverage within our maps, the SPT source
counts measured by \citet{vieira10} indicate we should
see $\simeq 2$ sources above 10~mJy in our maps.
Therefore, we detect approximately three times as many bright
radio sources as expected, although this excess is not
significant due to the small number of sources.

For the five clusters that contain these six radio sources
we have computed the SZ S/N both before and after
subtraction of the source(s).
Note that in all cases we have removed the bright central
radio sources described in Table~\ref{tab:radio_2}.
These six radio sources are located at a range of positions relative
to the cluster centers (between 2.5 and 6.6~arcmin),
and have 140~GHz flux densities between 8.7 and 51.4~mJy.
In most cases the SZ S/N decreases after subtraction
of the radio source(s), due to the fact that the
radio source(s) are mainly in the ring of positive flux
surrounding the SZ decrement caused by the high-pass
filtering applied to the data.
In the extreme case of Abell 963, with a 26.8~mJy radio
source located only 2.5~arcmin from the cluster center,
the fractional change in the SZ S/N is 12\%.
However, the median fractional change in SZ S/N due
to these bright sources is $\simeq 2.5$\%.
If we therefore assume that a source brighter than 2.5 times
below our detection threshold would be required to 
produce a $>1$\% fractional change in the SZ S/N,
then based on the results of \citet{vieira10} we expect
$2-3$ such sources among our sample of 45 clusters.
Therefore, given the results of Section~\ref{sec:central},
where 13 clusters had central radio galaxies that
produced a $>1$\% change in the SZ S/N,
contamination from undetected back/foreground
radio galaxies should be a factor of $\simeq 5$ below
the contamination from central cluster member radio galaxies.

After removing the six sources with NVSS associations, 
we are left with nine
unassociated pixels (six positive and three negative).
All of these unassociated pixels have S/N just above four.
Statistically, based on Gaussian noise,
we only expect $\simeq 1$ noise excursion
with S/N larger than four, and it is unclear why we have
nine such excursions.
However, we do note that most of our images have 
significant SZ signal in them (S/N~$>10$),
and relatively modest differences between our SZ model and
the underlying cluster profile could therefore combine with
noise excursions in some clusters and cause
such high S/N pixels in our best-fit-SZ-model-subtracted images.
In addition, the noise in our images is not perfectly white, and
large-angular-scale noise could also be responsible for some of these
S/N~$>4$ pixels. 
Given that our data show an excess of bright radio sources exclusive 
of these nine unassociated pixels, we assume that they are the
result of one of the non-idealities mentioned above rather
than an actual radio source.
We do not attempt to further account for these nine unassociated pixels
with S/N $>4$.

\section{Conclusions}

Using a combination of 140~GHz Bolocam data, 1.4~GHz NVSS data,
and $\simeq 30$~GHz data from OVRO/BIMA and SZA, we have 
studied the 140~GHz emission from cluster-member galaxies
in a sample of 45 massive clusters.
In agreement with previous results, we find that cool-core
clusters preferentially host more radio galaxies
and that all of the cluster-member galaxies have steep spectra.
On average, we find that the spectral index between
$30-140$~GHz is consistent with, but slightly steeper than,
the spectral index between $1.4-30$~GHz.
We further find that galaxy to galaxy variations lead
to a $\simeq 30$\% scatter in 140~GHz flux densities
extrapolated from data below 30~GHz.
Based on an extrapolation to 140~GHz from the lower frequency data,
we find that only $\simeq 1/4$ of the clusters contain
enough radio emission to produce more than a 1\% bias
in the optimally filtered peak SZ signal from the cluster, and the maximum
contamination in a single cluster is $\simeq 20$\%.
However, we do note that
the contamination in cool-core systems is significantly
enhanced compared to the sample average.
These results roughly match expectations from lower-frequency measurements
and simulations \citep{lin09, sehgal10, andersson11}
and indicate that the amount of radio contamination in high mass clusters
is small compared to achieved noise levels in SZ measurements obtained at 140~GHz.

\section{Acknowledgments}

We acknowledge the assistance of: 
the day crew and Hilo
staff of the Caltech Submillimeter Observatory, who provided
invaluable assistance during data-taking for this
data set; 
Mike Zemcov, Dan Marrone, and John Carlstrom for useful discussions;
Max Bonamente, John Carlstrom, Thomas Culverhouse, Christopher Greer, Marshall Joy, 
James Lamb, Erik Leitch, Dan Marrone, 
Amber Miller, Thomas Plagge, Matthew Sharp, and David Woody for providing
OVRO/BIMA and/or SZA data for our analysis;
Kathy Deniston, Barbara Wertz, and Diana Bisel, who provided effective
administrative support at Caltech and in Hilo;
Matt Hollister and Matt Ferry, who assisted in the
collection of these data;
the referee for useful suggestions that significantly improved
our manuscript.
The Bolocam observations were supported by the Gordon and Betty
Moore Foundation.
JS was supported by a NASA Graduate Student Research Fellowship,
a NASA Postdoctoral Program Fellowship, NSF/AST-0838261
and NASA/NNX11AB07G;
TM was supported by NASA through the Einstein Fellowship Program grant
PF0-110077;
NC was partially supported by a NASA Graduate Student
Research Fellowship;
AM was partially supported by NSF/AST-0838187;
SA, EP, and JAS were partially supported by NASA/NNX07AH59G;
KU acknowledges support from the Academia Sinica Career
Development Award.
A portion of this research was carried out at the Jet Propulsion
Laboratory, California Institute of Technology, under a contract
with the National Aeronautics and Space Administration.
This research made use of the Caltech Submillimeter Observatory,
which is operated by the California Institute of Technology
under cooperative agreement with the National Science Foundation 
(NSF/AST-0838261).
The operation of the SZA was supported by NSF/AST-0838187, 
and CARMA operations were supported by the CARMA partner universities
under a cooperative agreement with the National Science Foundation.

\end{document}